\documentclass{article}

\usepackage{PRIMEarxiv}
\usepackage{float}

\usepackage[utf8]{inputenc} 
\usepackage[T1]{fontenc}    
\usepackage{hyperref}       
\usepackage{url}            
\usepackage{booktabs}       
\usepackage{amsfonts}       
\usepackage{nicefrac}       
\usepackage{microtype}      
\usepackage{lipsum}
\usepackage{fancyhdr}       
\usepackage{graphicx}       
\graphicspath{{media/}}     

\usepackage[numbers]{natbib}
\setcitestyle{square}

\title{Learning a General Model of Single Phase Flow in Complex 3D Porous Media

}

\author{
  \textbf{Javier E. Santos} \\
  Los Alamos National Laboratory \\
  \And
  \textbf{Agnese Marcato} \\
 Los Alamos National Laboratory\\
  \And
  \textbf{Qinjun Kang} \\
 Los Alamos National Laboratory \\
  \And
  \textbf{Mohamed Mehana} \\
 Los Alamos National Laboratory\\
  \And
  \textbf{Daniel O'Malley} \\
 Los Alamos National Laboratory \\
  \And
  \textbf{Hari Viswanathan} \\
 Los Alamos National Laboratory \\
  \And
  \textbf{Nicholas Lubbers} \\
 Los Alamos National Laboratory \\ 
}

\begin{document}
\maketitle

\begin{abstract}
Modeling effective transport properties of 3D porous media, such as permeability, at multiple scales is challenging as a result of the combined complexity of the  pore structures and fluid physics - in particular, confinement effects which vary across the nanoscale to the microscale. While numerical simulation is possible, the computational cost is prohibitive for realistic domains, which are large and complex.  Although machine learning models have been proposed to circumvent simulation, none so far has simultaneously accounted for heterogeneous 3D structures, fluid confinement effects, and multiple simulation resolutions. By utilizing numerous computer science techniques to improve the scalability of training, we have for the first time developed a general flow model that accounts for the pore-structure and corresponding physical phenomena at scales from Angstrom to the micrometer. Using synthetic computational domains for training, our machine learning model exhibits strong performance (R$^2$=0.9) when tested on extremely diverse real domains at multiple scales.
\end{abstract}

\section{Introduction}

Estimating the transport properties of materials is challenging. On one hand, laboratory experiments are costly and require specialized facilities and in-hand processed samples that can not be re-used. On the other hand, numerical simulations often require complex meshing, assumptions and approximations, and are computationally intensive, especially for large complex domains. Alternatively, machine learning (ML) models are demonstrated to learn complex relations from large datasets with ease \cite{LeCun2015DeepLearning}. Most importantly, thanks to advances in software and hardware, ML models are able to provide predictions in new domains in milliseconds. This makes them attractive for forecasting physical processes, especially in the geosciences, where real-time predictions can be critical in field operations. However, training and testing machine learning models with subsurface data has many challenges. First, ML models are commonly optimized to 2D image benchmarks \cite{deng2009imagenet, Krizhevsky2009LearningImages}. On the other hand, imaging technologies (CT, FIB-SEM, MRI) have had breakthroughs \cite{Wildenschild2013X-raySystems} that allow us to obtain high definition samples that can be of size 256$^3$ or more; 3D domains are much more computationally challenging than most 2D images. Secondly, subsurface formations exhibit a wide range of complex structures due to the diverse rock genesis and diagenesis processes \cite{diagenesis}. Finally, it is not straightforward to embed the complexity of varying degrees of confinement effects in the ML model.  While much progress has been made in emulating transport in porous media using ML \cite{Marcato2022FromMedia,Kashefi2021Point-cloudPrediction,Wang2021ML-LBM:Networks,Ting2022UsingFractures,Cawte2022AData,Caglar2022DeepMicrostructures,Zhang2022PermeabilityNetwork,Tang2022PredictingInformation,Chang2022ElRock-Net:Simulations}, models in the literature are not general enough to be used as a generic tool.

Fluid transport driven by an external force, for example, a pressure gradient or the gravitational force, is affected by viscous, slip, and diffusive effects depending on the characteristic length-scale of the available flow paths. The primary way to quantify the impact of these effects on the ability of a fluid to travel through a sample is the \textit{apparent permeability}. This quantity provides a volume-average scalar that summarizes all the micro- and nano-scale effects into a single number, which can then be used in continuum simulators or analytical approximations for forecasting field-scale processes. A strict lower bound of the apparent permeability is the \textit{absolute permeability}, defined as the limit of the apparent permeability when nanoconfinement effects vanish. Slip and diffusive effects can enhance apparent permeability up to three orders of magnitude compared to absolute permeability, depending on the geometry and pressure (as described in \cite{Yin2022IdentifyingMedia} and shown in Figure \ref{fig:regimes}).

The structure and connectivity of the pore structures have the leading impact on the apparent permeability of a 3D porous medium. In micron-sized pores, fluid transport is approximated by \textit{Darcy's Law} where viscous forces are dominant with negligible contributions from inertial forces. During viscous flow, the velocity profile of the fluid is non-uniform and affected by the pore morphology and pressure gradient. This non-uniformity stems from the viscous drag forces which are uneven across the flow paths. Fluids prefer the least-drag force paths. For a tube with a circular cross-section, the viscous forces are minimal at the center of the pore given rise to a parabolic-shaped velocity profiles. In contrast, in smaller pores, transport is affected by diffusion and slip at the solid-fluid boundary since the mean free path length of the molecule is comparable to or larger than the pore size, increasing molecular collisions \cite{Javadpour2007} and enhancing permeability  \cite{Klinkenberg1941TheGases, Javadpour2009NanoporesSiltstone, Song2016ApparentReservoir}. In nature, permeable media hosts conduits and heterogeneities in a wide range of length-scales. Subsurface formations have pores that can be a few Angstroms wide \cite{Landry2016DirectPermeability, Loucks2012SpectrumPores} to fractures with apertures on the centimeter scale \cite{Marrett1999ExtentRock}. These wide range of scales encompass flow regimes going from free-molecular to viscous dominated \cite{Javadpour2007}. These  multiscale heterogeneities usually co-exist. For example, in fractured shale a fluid can flow much faster through the fracture, but the nanoporous matrix has a non-negligible contribution as shown by \cite{Chen2015GeneralizedEffect}.

Laboratory experiments can be set-up to study deviations from Darcy's law. However, since the resolution of the instruments is on the same scale as the phenomenon under investigation, running an experiment at subsurface conditions is very challenging. In practice, samples are subject to extremely high pressure gradients \cite{Tison1993ExperimentalLeaks, Marino2009ExperimentsTubes, Anovitz2015CharacterizationStructures,Curtis2012MicrostructuralImaging,Sondergeld2010Micro-StructuralShales} but this does not reflect the conditions that subsurface flow modellers are interested in. On the computational side, analytical solutions considering fluid flow inside a tube have been derived either from empirical observations or classical fluid dynamics and can be used with pore-network models \cite{Mehmani2013MultiscaleFlows}. Although these pore-network models try to capture the 3D connectivity of real samples, they aggressively oversimplify the heterogeneity of the pore space by  fitting it with spheres and tubes.  The high aspect ratio of the tubes will significantly influence the solution of non-Darcy flows and as seen in SEM images of mudrocks, pore structures vary from ``spongey'' kerogen patches, to long and skinny microfractures, to stacks of platelets \cite{xu_pore_2020}. Finally, molecular dynamics simulations can represent nanoconfinement effects accurately.  Nonetheless, the computational resources needed are vast, which limits the calculations to very small volumes (boxes with side lengths of around 10 nm).

The lattice-Boltzmann method (LBM) can operate in larger domains, which yield in representative sample sizes since it operates at a mesoscale. LBM integrates physical insights from both the molecular- and micron-scales and can predict larger-scale properties. In short, fluids are simulated as swarms of particles, represented by particle distribution functions, which flow on a discrete lattice. Landry \cite{Landry2016DirectPermeability} proposed the local effective viscosity model (LEV-LBM) to simulate multiple scales encompassing multiple flow regimes by utilizing a spatially-varying mean free path (sMFP). The sMFP accounts for the reduction of the local kinetic viscosity caused by the confinement effect at the nanoscale. One thing that sets apart the LBM simulation of that at the nanoscale, is that flow fields do not rely exclusively on the pore-structure but also on length scale, confinement pressure, temperature, and gas properties. The LEV-LBM also accounts for slip and diffusive effects depending on the scale. The main downside is that simulating a large complex sample with confinement effects could take weeks running on 1,000 processors.

While the LBM offers detailed insights into fluid dynamics, there is a growing interest in leveraging this data and the computational efficiency of neural networks for permeability modeling. Neural networks are an attractive avenue for modeling permeability since they can provide predictions in a split-second enabling real-time field applications. Therefore, there have been many attempts to train neural networks to predict permeability from images. In a pioneering study, \cite{Wu2018SeeingNetworks} employed a convolutional network (ConvNet) to learn a mapping between a 2D image and its permeability, since then ConvNets have remained the model of choice for this task. \cite{Wang2021DeepModeling} provides a comprehensive review of notable works to this date. Nevertheless, computational scaling with domain size is still one of the main hurdles in training models with 3D images of permeable media. Recently, we provided a proof of concept of a network that allowed training and testing with large 3D images on a single graphics processing unit, capturing representative elementary volumes  within the weights of the network \cite{Santos2021ComputationallyMedia}.

In this letter,  we present a \textit{general model for fluid transport through permeable media} building upon the work of MS-Net \cite{Santos2021ComputationallyMedia} by improving the model design, scaling-up the parameter count (model size), and increasing the size and diversity of the training set. For the first time, we used hundreds of computationally large flow simulations in synthetic samples to train our model, including a wide range of pore sizes, rock types, and boundary conditions. The comprehensive dataset used for all the training and testing data are publicly available \cite{Santos2022AMedia}. To accommodate the larger model and dataset, we likewise applied a number of computer science tools to ameliorate the increased computational cost, which we detail. Our contributions are divided into two main categories, 1) model architecture, where we the describe in detail the model that allow for multi-scale processes to be accounted for and 2) training, which is essential for distilling information from such a diverse set of samples properly and its vital for managing a model of this size and complexity. This workflow is general and transferable to other properties and domains. Finally, we show that the trained model is able to generalize the problem of fluid flow through permeable media for more than one thousand simulations, including many microCT images of real porous media.

\begin{figure}[H]
    \centering
    \includegraphics[width=1.1\textwidth]{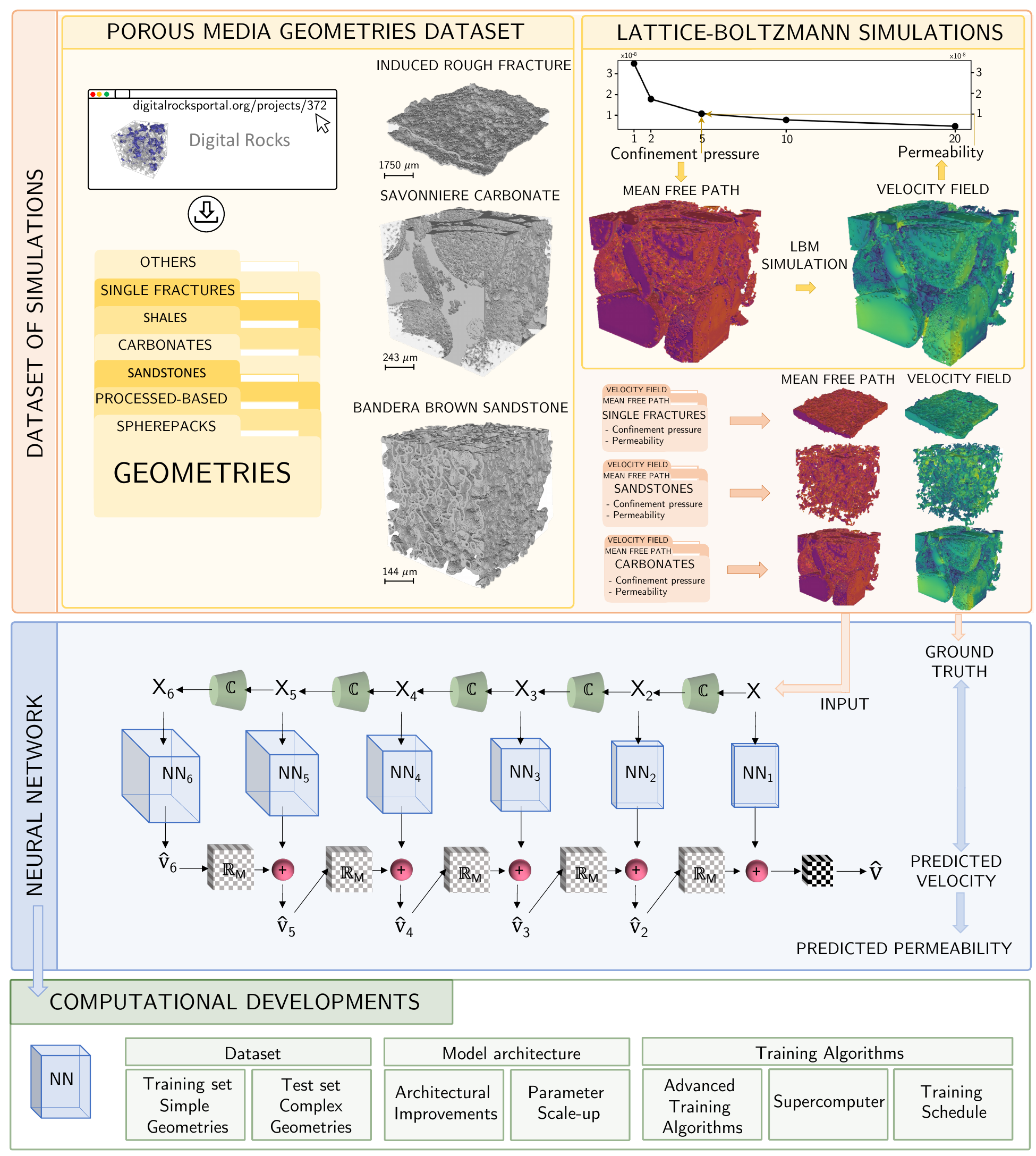}
    \caption{Workflow figure. 
    }
    \label{fig:workflow}
\end{figure}

\section{Machine learning workflow}

We describe the machine learning workflow in the following subsections. Dataset describes the data used for both training and testing. Model architecture describes the functional form of the machine learning model. Training algorithms explains the various techniques required to scale the training to large datasets and complex geometries, including the training schedule which describes the actual train procedure. An schematic showing all the stages of this process is shown in Figure \ref{fig:workflow}.

\subsection{Dataset}

\begin{figure}
    \centering
    \includegraphics[width=1.\textwidth]{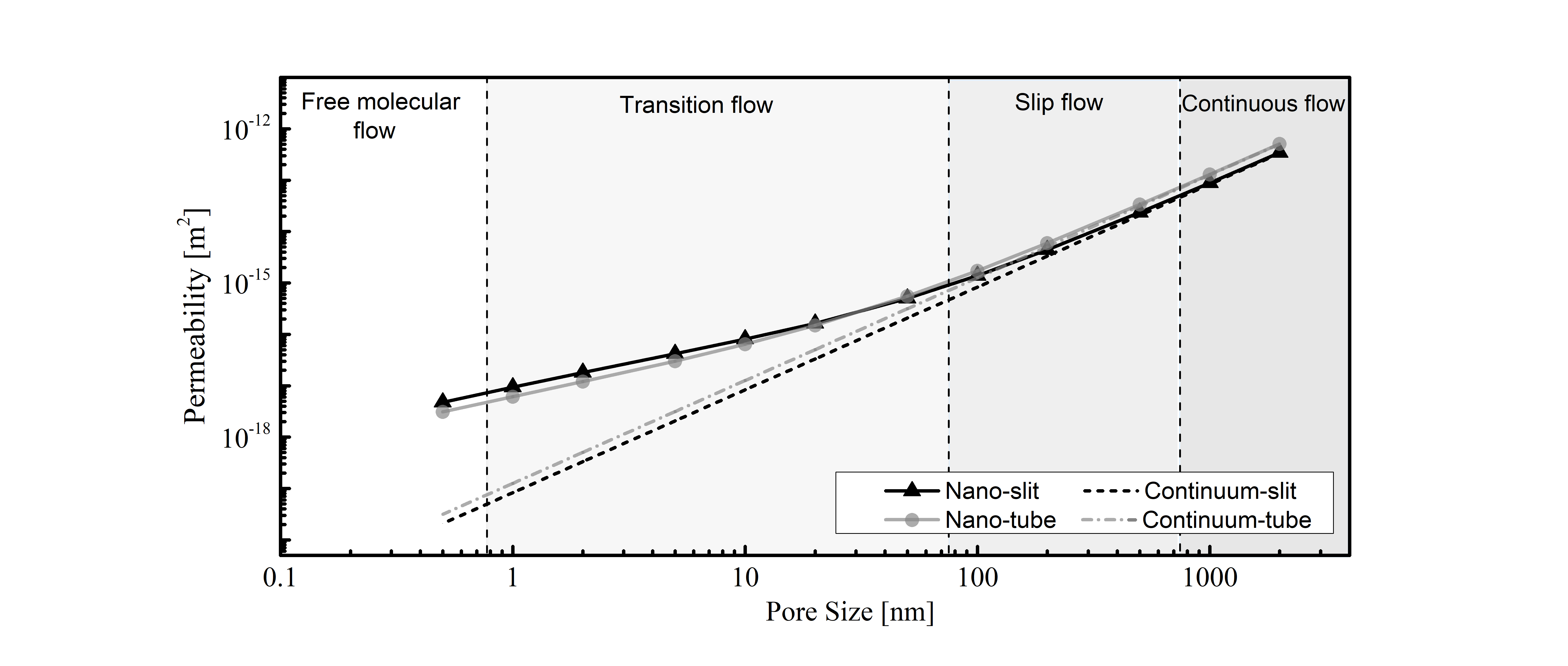}
    \caption{Flow regimes for simplified samples, in reality, these effects are magnified due to the complexity of the 3D structures, for example, how tortuous the paths are of how narrow some paths can get.}
    \label{fig:regimes}
\end{figure}

For a thorough review of the dataset, we refer the reader to \cite{Santos2022AMedia}. The  DRP-372 dataset includes more than 1200 flow simulations from a wide variety of samples hosted in the Digital Rocks Portal \cite{Prodanovic2015DigitalImages}. We categorized the dataset based on 6 lithology groups: spherepacks, processed-based geometries, shales, carbonates, sandstones, and single fractures. The flow simulations were carried-out in computational sizes of 256$^3$ and 480$^3$ where some of the latter took up to 4 days running on 1,000 CPU cores to complete. The resolution of these samples spanned from 0.5 nanometers to 5 micrometers. These simulations were carried-out using a local-effective viscosity lattice-Boltzmann model \cite{Landry2016DirectPermeability} which captures the impact of  nanoconfinement effects, slip boundary conditions, and diffuse boundary layers when applicable. The confinement pressure was varied among individual simulations yielding a dataset that captures a wide range of scales and flow regimes (Figure \ref{fig:regimes}). The lower pressure bound is set at 1 MegaPascal (MPa) where significant nanoconfinement effects are present. The upper pressure was 20 MPa, where negligible confinement effects were shown and the solution is the Navier-Stokes equation solution. These simulations encompass flow regimes from the free molecular to viscous flow. The data spans 9 orders of magnitude of apparent permeability. The computational cost of these simulations required more than 5,000,000 core-hours.

To train our model, we utilized 60 simple synthetic geometries described in \cite{Santos20223DMedia}, including various types of sphere packs, both homogeneous and heterogeneous, and fractures. These process-based samples are designed to emulate various lithologies. These synthetic samples feature well-resolved pores and interconnected structures in the flow direction. The main motivation to use synthetic samples is to show that our model is able to abstract how fluid flow behaves under different physical conditions and is able to be used in a wide range of unseen real geometries. 

Consequently, during testing, we evaluate the model's performance predominantly on real porous media, thereby highlighting its capacity to transfer knowledge from synthetic to real-world scenarios. This dataset includes 75 diverse samples from research groups worldwide \cite{bultreys_belgian_2020, bultreys_estaillades_2018, van_offenwert_saturated_2019, spurin_decane_2021, armstrong_moura_2017, singh_high_2018}, captured using various imaging devices or simulated using different process-based codes. By leveraging this extensive dataset, our study rigorously evaluates the generalization capabilities of our model across a wide range of real-world samples.

\begin{figure}
    \centering
    \includegraphics[width=1.\textwidth]{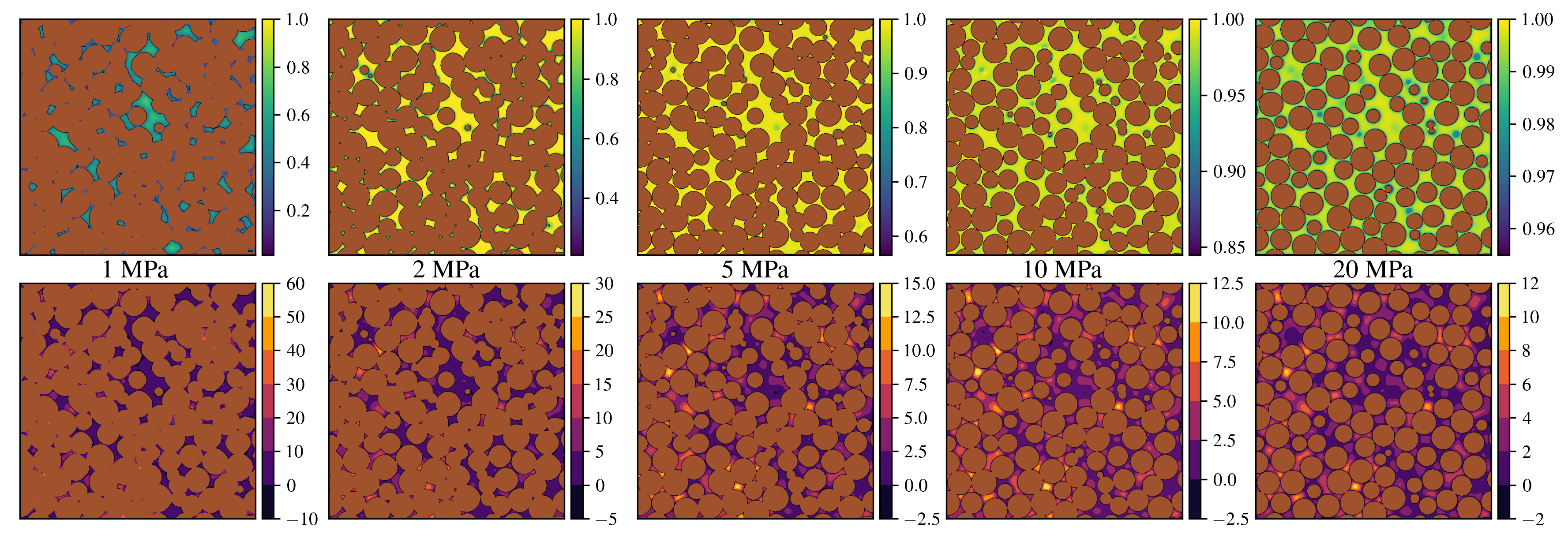}
    \caption{The bottom panel shows a cross-section of the nanoconfined simulation at five different pressures. The normalized velocity (with respect to the maximum value) contours are shown. It is visible that the sample at the lowest pressure (1 MPa) presents the higher degree of slip, shown as “fat“ velocity profile (fewer contour lines), while the one at 20 MPa exhibits no slip. This can also be seen in the top panels, where normalized mean free path (that could be interpreted as the deviation from the Navier-Stokes solution) is shown.}
    \label{fig:mfp}
\end{figure}

\subsection{Model architecture}

Traditionally, analytical solutions have been proposed for permeability estimation using average (effective) structural properties, the most prominent example is the use of the porosity. Nevertheless, the average porosity of a sample is a non-unique proxy (many samples have the same average porosity value) and fails to capture the sample's pore-space complexity and spatial distribution. Another notable example is the fracture width used in the cubic law to estimate single-fracture permeability. To get a more complete picture, our proposed ML model uses the whole complex 3D structure, with its thermodynamic conditions and scale. These are all included to aid the network to compute the resultant apparent permeability under these conditions.  We explain in detail our network architecture in the following subsection.

We build our model using a larger version of the MS-Net \cite{Santos2021ComputationallyMedia} architecture, which demonstrated a remarkable ability to learn from large porous media samples. In short, the MS-Net is a system of convolutional networks that process data at different resolutions. The system is trained end-to-end by passing the outputs of the coarser networks sequentially as inputs to the finer ones \footnote{The details about the coarsening and refinement procedure can be found in Section 2.4 of \cite{Santos2021ComputationallyMedia}.}. There are two main motivating factors for using this architecture: 1) This is an elegant way to process large  ($>$256$^3$) geometries without subsampling the domain, and 2) the coarsest model can see the entire image at once and predict the bulk of the flow while the subsequent ones refine it. In this work, we have implemented the following enhancements to the previous model:

\begin{itemize}
    \item \textbf{Input features}: To capture simulations encompassing diverse thermodynamic conditions, we introduce a spatially-varying 3D normalized mean free path in our model. This feature, which arises from the Peng-Robinson equation, accounts for the reduction of the local kinetic viscosity caused by the confinement effects at the nanoscale (the detailed explanation can be found in  \cite{Santos2022AMedia})\footnote{Initially, we tried training our network with only this input. The trained model provided accurate predictions of permeability but the flow was not spatially-smooth, since this map does not have enough variance, as shown in Figure \ref{fig:mfp}}. We also include the 3D Euclidean distance of the pore-space. The Euclidean distance labels each voxel with the distance to the closest solid boundary.\footnote{The Euclidean distance outputs large values in the middle of the pores and smaller values near the solid boundaries, is amenable to the cylindrical profile of a fluid flowing on laminar flow.} This feature aids to get a spatially-smooth prediction. The computational cost of these two features is negligible, even in very large domains.

    \item \textbf{Large number of scales}: To significantly expand the model's field of vision, we augmented our system by incorporating six convolutional sub-networks. This design decision enables the coarsest network (NN$_6$ in Figure \ref{fig:workflow}) to receive an input that is $\frac{1}{32}$th the size of the original domain. Despite the downsizing, the input retains crucial global structural features that affect the fluid flow, which allows the kernels in the sixth network to see most of the image at once.

    \item \textbf{Padding}: To facilitate the model's comprehension of flow continuity along the flow direction, we incorporated periodic padding for the 3D convolution operations.  Meanwhile, since the faces perpendicular to this axis feature no-flow boundaries in the simulation, we assigned zero-padding to these regions. These  design choices are particularly valuable in the networks operating at coarser scales, as their field of vision is comparable to the input size (At the coarsest scale, a 256$^3$ domain corresponds has an input size of 8$^3$).

    \item \textbf{No flow in the solid-space}: To explicitly ensure no flow within the solid space, we introduced an additional constraint at the end of our model. The final output of the model is multiplied by the input binary image, which contains ones in the pore-space and zeros in the solid-space. This multiplication effectively disregards the gradients of voxels outside the pores. By significantly reducing the search-space, this approach leads to notably faster training. Additionally, we conducted experiments where we allowed gradients of the first solid voxel surrounding the grains, but observed no discernible difference in the training process.

    \item \textbf{Fewer activation functions and normalization layers}: To enhance accuracy, we implemented a strategy recommended in \cite{Liu2022A2020s} by reducing the number of activation functions and normalization layers in our model. Specifically, we removed the normalization layer preceding the last convolutional layer, as well as the activation layer following it. This adjustment resulted in an improvement in accuracy.

    \item \textbf{GeLU activation}: To align with recent advancements in models, we transitioned our remaining activations to Gaussian Error Linear Units (GeLU) \cite{Hendrycks2016GaussianGELUs}. GeLU serves as a smoother alternative to ReLU and has gained popularity in the field. By adopting GeLU, we achieved a slight increase in accuracy while maintaining the same computational overhead.

    \item  \textbf{Number of filters at each layer}: Initially, we experimented with increasing the depth of each network to enhance model capacity. However, this approach proved challenging to train, likely due to increasingly difficult gradient paths. Consequently, we opted for a substantial increase in the number of parameters, significantly raising the count to 400M compared to the 10k of the MS-Net. This substantial capacity boost resulted in a significant improvement in accuracy. Notably, the augmented model's training process required adjustments to the training algorithm, as elaborated in Section \ref{sec:algos}, to effectively handle the increased memory requirements.

\end{itemize}

In addition, we explored the inclusion of attention layers in the coarser networks. While this design choice led to slight improvements in accuracy, it also incurred significant computational costs. Furthermore, incorporating attention mechanisms posed challenges in accommodating different image sizes, making it less practical for our purposes. As a result, we opted for a fully convolutional architecture, which better addressed the need for flexibility in working with diverse image sizes.

\subsection{Training algorithms}
\label{sec:algos}

The development of an enhanced model, encompassing larger networks with increased parameter counts, and training it on computationally larger domains with more features, has yielded a notable improvement in accuracy for both training and testing datasets. However, scaling up the model presented significant challenges, particularly in addressing the memory constraints imposed by the GPUs. In this section, we emphasize the algorithms that were crucial in overcoming these obstacles and enabling the successful implementation of the model.

\begin{itemize}
 
    \item \textbf{Mixed-precision training}: In order to accommodate the larger model and facilitate training with bigger domains, we carried out the training process using 16-bit precision. This approach offers several advantages, including reduced memory consumption as the network parameters occupy half the space, and accelerated data transfer due to operations requiring less bandwidth and executing at a faster pace. However, it is important to note that training stability can be compromised due to precision loss, which poses a significant challenge in our specific case.

    \item \textbf{Activation checkpointing}: To reduce GPU memory usage, we implemented activation checkpointing in our model. This technique avoids the storage of layer activations, recomputing them during the backward pass instead. By sacrificing some computation time, we achieve a reduction in memory requirements, allowing us to train larger models with the same memory footprint.

    \item \textbf{Gradient accumulation}: Even with the utilization of 16-bit precision, the size of our model and samples restricts us from performing a backward pass with more than one domain at a time. Since training with a single sample per step (\textit{online training}) would be extremely noisy, we use gradient accumulation to ``simulate'' bigger batches. The idea behind gradient accumulation is simple: after calculating the loss and gradients for each sample, instead of immediately updating the model parameters, the gradients are accumulated across consecutive samples. The model parameters are then updated based on the cumulative gradient after a specified number of samples. This approach serves the same purpose as having a multiple sample mini-batch, all while maintaining the memory footprint of a single sample.

     \item \textbf{Loss function}: Our dataset encompasses a wide range of scales, with permeabilities spanning 8 orders of magnitude and individual pixel-wise velocities spanning 24. To effectively convey the relative importance of each sample to the model, we employed a loss function that calculates the pixel-wise error relative to the variance of each velocity field. This approach ensures that every sample carries the same level of "importance" during training.
     
\begin{equation}
    \mathcal{L} = \sum_{i=1}^{g_{steps}}  \sum_{s=1}^{6} \frac{ {\rm sum}((u_{i,s}-\hat{u}_{i,s})^2)}{N_{vox,s} \cdot \sigma(u_{i})^2}. 
\end{equation}

 $g_{steps}$ are the number of gradient accumulation steps, $u_{i}$ represent each of the 6 coarsened velocity fields, $N_{vox,s}$ the number of voxels per scale and $\sigma(u_{i})^2$ the standard deviation of the velocity field. Our loss function tries to reconstruct this. Solely computing the mean squared error in the last layer would assign greater importance to samples with higher velocities.

    \item \textbf{Input and output normalization}: To enhance training speed and stability, we normalized the Euclidean distance with a constant value of 50 and the velocity with a value of 1e-9. These value choices accelerated the training process and enhanced its stability. Additionally, we kept the normalized mean free path in its original range of zero to one without further normalization.

    \item \textbf{Gradient clipping}: To mitigate potential instabilities during training, we employed gradient clipping, a technique that limits the magnitude of gradients to a specified value. By rescaling the error derivative, the weight updates are also appropriately adjusted, minimizing the risk of overflow or underflow. This approach ensures that gradient descent exhibits stable behavior, even when faced with a complex and irregular loss landscape. By implementing gradient clipping, we successfully prevent training instabilities from adversely affecting the model's training path.

   \item \textbf{Optimizer}: To enhance training stability and improve generalization to unseen data, we utilized the AdamW optimizer during model training. One key benefit of AdamW is the decoupling of weight decay from gradient updates, which encourages the model to maintain smaller weights, promoting stability throughout the training process and generalizing to unseen data better. 

\end{itemize}

\subsubsection{Training schedule}

Given the significantly increased parameter count resulting from the strategies detailed in the previous section, combined with the diversity of the dataset, and the inherent instability of 16-bit training, network initialization plays a critical role. However, when using the weight initialization strategy proposed in \cite{Santos2021ComputationallyMedia}, we faced challenges due to the wide range of orders of magnitude in the individual pixel-wise velocity samples, which lead to overflow or underflow issues shortly after a few training iterations. To address this issue, we devised a solution that involved "warming up" the model weights using the simplest geometry, spherepacks. To ensure the model's weight expressiveness in predicting a wide range of velocity outputs, we trained the model using five 256$^3$ spherepacks. These geometries covered porosity ranges of 8 to 25\% and pressures ranging from 1 to 20 Mpa (refer to Figure \ref{fig:mfp}). Although these spherepacks had similar permeability values, the velocities spanned a significantly broader range. This diverse training data enabled the model's weights to capture the necessary complexity and variability required for training with more diverse samples. This training process, conducted over 72 hours with a learning rate of 1e-6 and a batch size of one, necessitated the incorporation of a constraint. If the loss exceeded 1e-4, the corresponding sample was ignored, and the network parameters were not updated, since very high losses triggered network updates results in the training diverging. This constraint, enforced through the PyTorch distributed data parallel engine, effectively identified and froze the unused parameters. While this constraint was frequently triggered during the early stages of training, it gradually diminished over time, and disappearing completely after a few hours, allowing the training process to effectively adjust the weights.

After achieving stabilization of the loss function, we implemented a multi-stage training approach. Initially, we added one extra sample from each spherepack, selected randomly at a specific pressure. During this stage, we utilized an increased learning rate of 1e-5 and a batch size of two, which continued for 24 hours.  Subsequently, we included all the samples sized 256$^3$ in the training process, further increasing the learning rate to 1e-4 and expanded the batch size to 8 for an additional 24 hours. Finally, in the last training stage, we included the remaining samples (the 480$^3$ volumes) and performed eight random augmentations (including a 180-degree flip in the flow axis and 90, 180, 270-degree flips on the XY-axes). This resulted in approximately 1000 samples of varying sizes to train on. We carried out this process for approximately 168 hours. In the final stage, we fine-tune our model by incorporating stochastic weight averaging (SWA) to further enhance its performance. SWA offers notable advantages as it is easy to implement, improves generalization, and imposes minimal computational overhead. This aligns with the findings of \cite{Song2022BenefitsPrediction}, who also reported enhanced generalization capabilities through the use of SWA. Throughout this last stage, we utilized a 10\% validation split and saved the model with the best loss value observed in this set. To prevent overfitting, early stopping was employed after 300 epochs without improvement.

\section{Results}
Our primary focus is the apparent permeability of a 3D structure under varying confinement pressures. While it is possible to directly train a model to predict this quantity as a floating-point number, our preliminary experiments revealed a tendency for such models to overfit quickly and exhibit poor generalization to unseen data. Therefore, we trained our model using the complete flow field, even though our main interest lies only in the apparent permeability. In essence, the prediction of the flow field serves as an auxiliary task to support the training process. We would like to highlight the significant distinction in computational requirements between the creation of the training data and the testing phase. Generating the training data was accomplished with relative ease, requiring around 10,000 CPU hours due to the simplicity of the geometries. In contrast, the testing phase demanded a substantial investment of computational resources, totaling close to 5 million hours of CPU time. 

\begin{figure}
    \centering
    \includegraphics[width=1.1\textwidth]{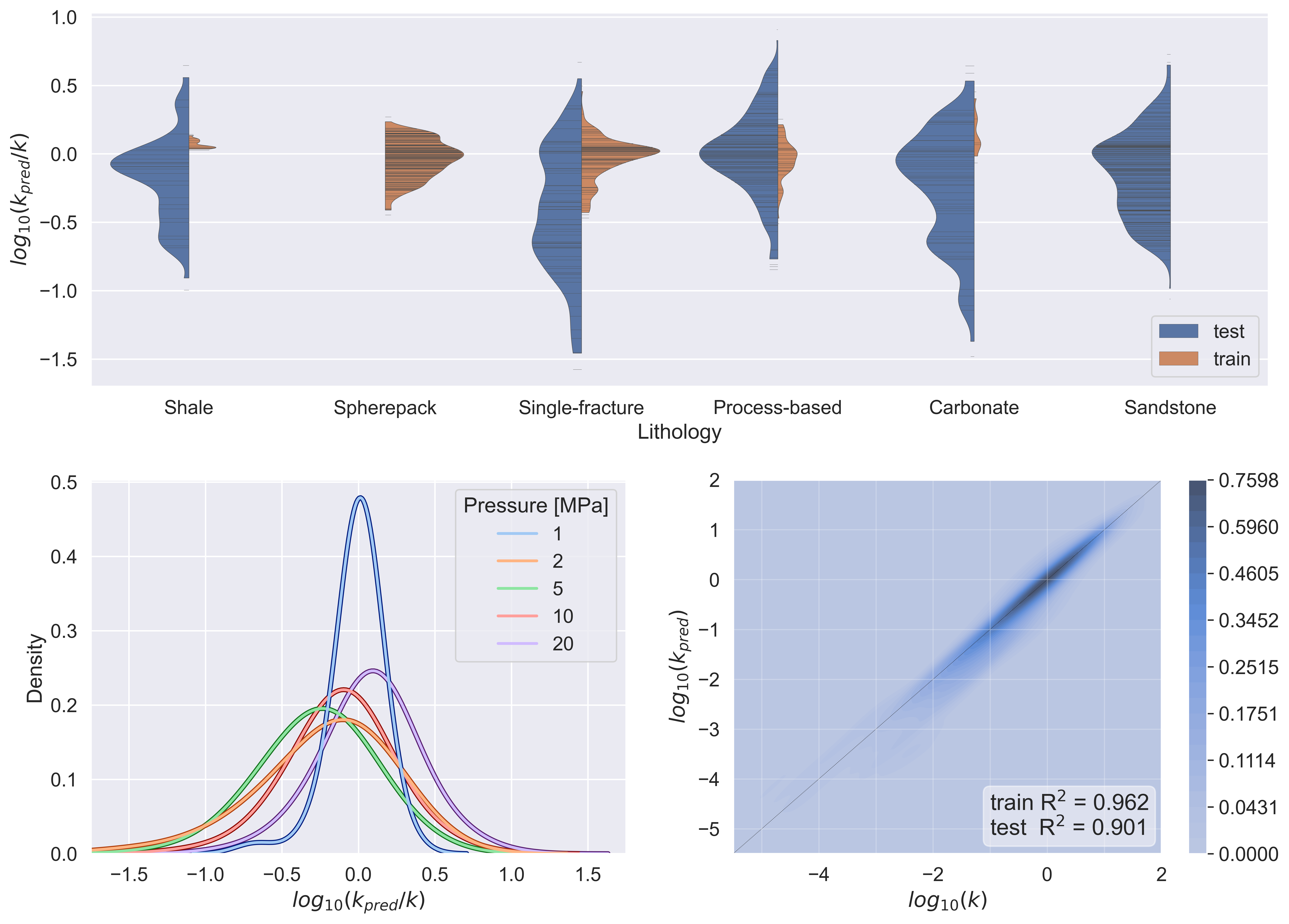}
    \caption{\textbf{a)} Performance of the model in the training and testing sets for each lithology. Each black bar represents one sample. \textbf{b)}  Performance of the model for different confinement pressures   \textbf{c)} Performance of the model in all the samples of the test set. The coefficients of determination for the train and test sets are shown in the lower right corner. }
    \label{fig:3D_results}
\end{figure}

We evaluated the performance of our model on unseen samples by quantifying the ratio between predicted and simulated apparent permeability (refer to Figure \ref{fig:3D_results}). The model's performance was assessed across different lithologies and pressures. Remarkably, the model demonstrated excellent performance on unseen samples, despite being exclusively trained on synthetic data. We observed a well-defined normal distribution for the process-based samples in the test set, which closely resemble the training set. For the remaining lithologies, we observed strong agreement between the predicted and simulated permeability values. Notably, the single fracture lithology exhibited the lowest performance, likely due to its distinct characteristics such as roughness, tortuosity, and pinch points, which differ from the single-fracture counterparts in the training set. Confinement effects are magnified as pore size and pressure decrease, we noted that our model tends to exhibit stronger agreement at lower confinement pressures, suggesting that the input feature of mean free path captures more salient features under such conditions. In terms of accuracy,most of the test samples demonstrate results within an order of magnitude. This achievement is particularly remarkable considering that the permeability of the test data spans eight orders of magnitude and local phenomena encompasses regimes from free molecular to viscous dominated.

\section{Conclusions}

For decades, approximations such as the cubic law in fractures, variations of the Kozeny-Carman equation for granular media, and empirical equations based on laboratory measurements have been relied upon to estimate permeability. However, these approaches are known to have significant limitations and inherent flaws. In contrast, machine learning provides a promising avenue for permeability estimation, offering enhanced accuracy and improved performance. Our developed model represents a significant advancement in this field, setting a new state-of-the-art benchmark for permeability estimation.

We have developed a data-driven model that utilizes a large and diverse dataset to accurately predict the apparent permeability of complex pore networks. This model can be easily fine-tuned with new data, enabling improved accuracy for specific types of porous media, or can be directly utilized as is. To improve performance and avoid overfitting, we carefully optimized the model size and training methods.  Our approach incorporates a comprehensive dataset, encompassing a wide range of sample sizes, features, and boundary conditions.

By employing a convolutional neural network, we trained our model on synthetic data and tested it on realistic heterogeneous samples, spanning various thermodynamic conditions, pore sizes, pore structures, and computational sample sizes. The versatility of this model lends itself to numerous practical and scientific applications. For instance, it can aid in establishing new relationships for fluid flow in porous media, and quantifying the effects of diagenetic processes over time. Since this model is extremely fast, The real-time prediction aspect could be exploited to optimize image segmentation parameters for binarization during x-ray scanning, estimate relative permeability curves while imagining an experiment, optimize porous media synthesis for specific properties, and permeability estimation from  cuttings to complement logging while drilling, among others.  In a broader view, our approach could be adopted for other phenomena and domains which will benefit from real-time predictions of commonly challenging-to-estimate properties.

\section{Open Research}
All the data used to train and test our model is published as \cite{Santos2022AMedia} and can be found at \url{https://www.digitalrocksportal.org/projects/372}.

\bibliographystyle{unsrt}

\end{document}